\documentclass[fleqn,usenatbib]{mnras}

\usepackage{newtxtext,newtxmath}
\usepackage[T1]{fontenc}

\DeclareRobustCommand{\VAN}[3]{#2}
\let\VANthebibliography\thebibliography
\def\thebibliography{\DeclareRobustCommand{\VAN}[3]{##3}\VANthebibliography}

\usepackage{graphicx}	% Including figure files
\usepackage{amsmath}	% Advanced maths commands
\usepackage{xcolor}
\usepackage{pdflscape}
\usepackage{orcidlink}
\usepackage{comment}
\usepackage{multirow}

\newcommand{\Msun}{$M_\odot$}
\newcommand{\Mstars}{$M_\star$}

\newcommand{\Ha}{H$\alpha$}
\newcommand{\OIIIHB}{{\sc [Oiii]}+H$\beta$}

\title[A merger of ultra-low-mass galaxies at $z\sim5$]{
JWST catches the assembly of a $z\sim5$ ultra-low-mass galaxy 
}

\author[Asada et al.]{
Yoshihisa Asada$^{1,2}$\thanks{e-mail: asada@kusastro.kyoto-u.ac.jp,  marcin.sawicki@smu.ca}\orcidlink{0000-0003-3983-5438},
Marcin Sawicki$^{1 \star}$\orcidlink{0000-0002-7712-7857},
Guillaume Desprez$^{1}$\orcidlink{0000-0001-8325-1742},
Roberto Abraham$^{3}$\orcidlink{0000-0002-4542-921X},
Maruša Bradač$^{4}$\orcidlink{0000-0001-5984-0395},\newauthor
Gabriel Brammer$^{5}$\orcidlink{0000-0003-2680-005X},
Anishya Harshan$^{4}$\orcidlink{0000-0001-9414-6382},
Kartheik Iyer$^{6,7}$\orcidlink{0000-0001-9298-3523},
Nicholas S. Martis$^{1,8}$\orcidlink{0000-0003-3243-9969},
Lamiya Mowla$^{6}$\orcidlink{0000-0002-8530-9765},\newauthor
Adam Muzzin$^{9}$\orcidlink{0000-0002-9330-9108},
Ga\"el Noirot$^{1}$,
Swara Ravindranath$^{10}$\orcidlink{0000-0002-5269-6527},
Ghassan T. E. Sarrouh$^{9}$\orcidlink{0000-0001-8830-2166},
Victoria Strait$^{11,5}$\orcidlink{0000-0002-6338-7295},\newauthor
Chris J. Willott$^{8}$\orcidlink{0000-0002-4201-7367},
Johannes Zabl$^{1}$\orcidlink{0000-0002-9842-6354}
\\
$^{1}$ Department of Astronomy \& Physics and Institute for Computational Astrophysics, Saint Mary's University, 923 Robie Street, Halifax, NS B3H 3C3, Canada\\
$^{2}$ Department of Astronomy, Kyoto University, Sakyo-ku, Kyoto 606-8502, Japan\\
$^{3}$ David A. Dunlap Department of Astronomy and Astrophysics, University of Toronto, 50 St. George Street, Toronto, Ontario, M5S 3H4, Canada\\
$^{4}$ University of Ljubljana, Department of Mathematics and Physics, Jadranska ulica 19, SI-1000 Ljubljana, Slovenia\\
$^{5}$ Niels Bohr Institute, University of Copenhagen, Jagtvej 128, DK-2200 Copenhagen N, Denmark\\
$^{6}$ Dunlap Institute for Astronomy and Astrophysics, 50 St. George Street, Toronto, Ontario, M5S 3H4, Canada\\
$^{7}$ Columbia Astrophysics Laboratory, Columbia University, 550 West 120th Street, New York, NY 10027, USA\\
$^{8}$ National Research Council of Canada, Herzberg Astronomy \& Astrophysics Research Centre, 5071 West Saanich Road, Victoria, BC, V9E 2E7, Canada\\
$^{9}$ Department of Physics and Astronomy, York University, 4700 Keele St. Toronto, Ontario, M3J 1P3, Canada\\
$^{10}$ Space Telescope Science Institute, 3700 San Martin Drive, Baltimore, Maryland 21218, USA\\
$^{11}$ Cosmic Dawn Center (DAWN), Denmark\\
}

\date{Accepted XXX. Received YYY; in original form ZZZ}

\pubyear{2023}

\begin{document}
\label{firstpage}
\pagerange{\pageref{firstpage}--\pageref{lastpage}}
\maketitle

\begin{abstract}
%\MNRASinstructions{This is a simple template for authors to write new MNRAS papers. The abstract should briefly describe the aims, methods, and main results of the paper.It should be a single paragraph not more than 250 words (200 words for Letters). No references should appear in the abstract.}\\
%$\MNRASinstructions{Max 200 words for a Letter}
\\
Using CANUCS imaging we found an apparent major merger of two $z\sim5$ ultra-low-mass galaxies (\Mstars~$\sim10^{7}$\Msun\ each) that are doubly imaged and magnified $\sim$12--15$\times$ by the lensing cluster MACS 0417. Both galaxies are experiencing young ($\sim$100~Myr), synchronised bursts of star formation with $\log({\rm sSFR/Gyr^{-1}}
)\sim$1.3-1.4, yet ${\rm SFRs}$ of just $\sim 0.2 M_\odot\ {\rm yr}^{-1}$. They have sub-solar ($Z\sim0.2Z_\odot$) gas-phase metallicities and are connected by an even more metal-poor star-forming bridge.  The galaxy that forms from the merger will have a mass of at least \Mstars~$\sim 2\times10^7$~\Msun, at least half of it  formed during the interaction-induced starburst. More than half of the ionizing photons produced by the system (before and during the merger) will have been produced during the burst. This system provides the first detailed look at a  merger involving two high-$z$ ultra-low-mass galaxies of the type believed to be responsible for reionizing the Universe. It suggests that such galaxies can grow via a combination of mass obtained through major mergers, merger-triggered starbursts, and long-term in-situ star formation. If such high-$z$ mergers are common, then merger-triggered starbursts could be significant contributors to the ionizing photon budget of the Universe. 
\end{abstract}


%%%%%%%%%%%%%%%%%%%%%%%%%%%%%%%%%%%%%%%%%%%%%%%%%%

%%%%%%%%%%%%%%%%% BODY OF PAPER %%%%%%%%%%%%%%%%%%

\begin{figure*}
	\includegraphics[width=2.08\columnwidth]{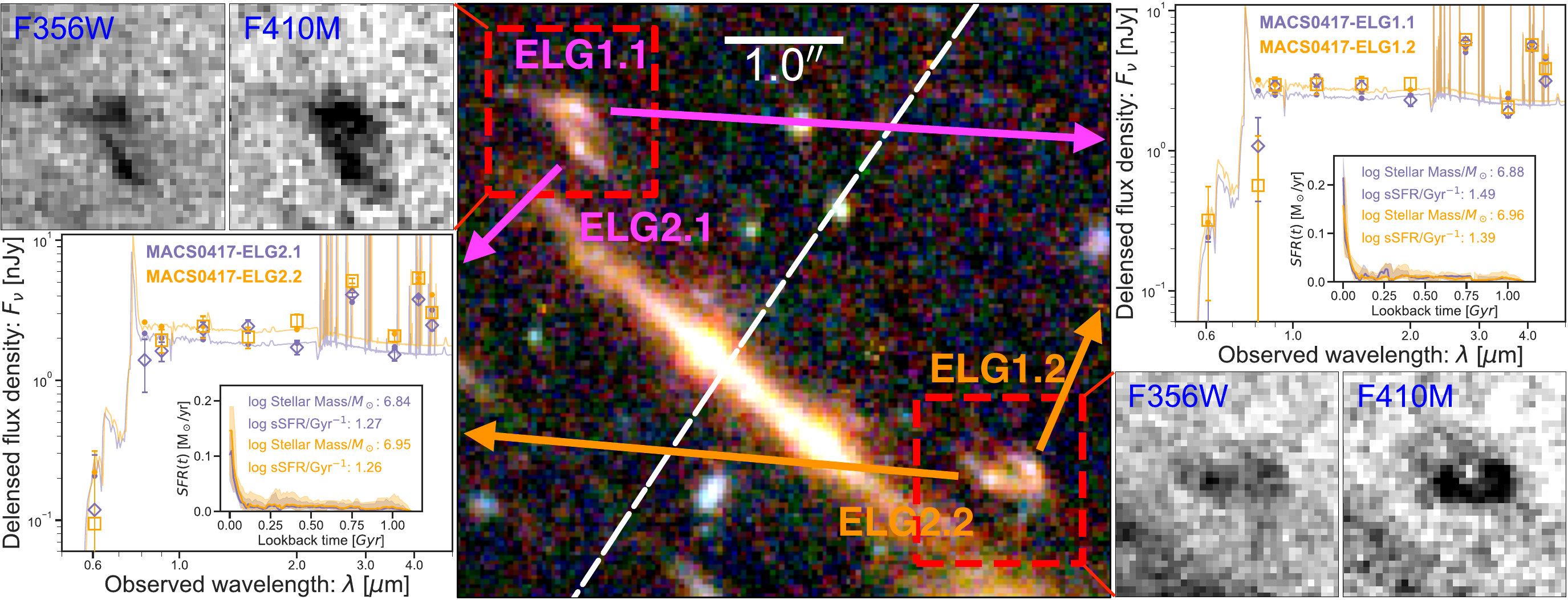}
    \caption{The centre panel shows the RGB composite image of the lensed system behind the galaxy cluster MACS J0417.5-1154.
    Magenta and orange show the north and south images of the galaxy pair, respectively.
    The white dashed line denotes the critical curve at $z\sim5.1$.
    The grayscale images in the left and right panel are direct images of the mergers (north and south) in F356W (continuum) and F410M (continuum + \Ha) with 1.4" on the side.
    The SED panels show observations (open symbols), the best-fiting model spectra (solid lines), and broadband SEDs synthesized from the models (filled circles). 
    The panels inset in the SED plots show the SFHs derived with {\tt dense basis} (see Sec.~\ref{subsec:sed} for details).
}
    \label{fig:RGB_SED}
\end{figure*}

\section{Introduction}

Given the low number densities of AGN and luminous galaxies at high redshift, it falls to low-luminosity galaxies to reionize the Universe \citep[e.g.,][]{Bouwens2015, Robertson2015, Iwata2022}. The census of such high-$z$ low-luminosity galaxies is thus a key goal that motivated the construction of JWST \citep{Gardner2023}, and while these surveys are still in their early days, we can already start using Webb to tackle questions about the nature of these objects.  One such question is ``how are the low-luminosity galaxies assembled at high redshift?''  

Since most high-$z$ galaxies are star-forming, an obvious growth channel is through steady star formation (SF). On the other hand, recent works suggest bursty SF is common in the high-$z$ universe and can play an important role in galaxy evolution \citep{faisst_recent_2019,Rinaldi_2022}. However, particularly for low-mass galaxies at high-$z$, the mechanism that triggers bursty SF is as yet unclear. Studies of local star-bursting dwarfs \citep[e.g.,][]{Lelli2014} suggest that an external mechanism, including galaxy interactions/mergers, can trigger bursts of SF in low-mass galaxies, but strong evidence for low-mass high-$z$ merging galaxies experiencing SF bursts has not yet been reported.
Thus, finding examples of low-mass galaxies in the high-$z$ universe undergoing bursts of SF induced by galaxy-galaxy interactions is important: it can demonstrate such an external mechanism can trigger SF bursts in high-$z$ low-mass galaxies and enable us to study the properties and impact of such bursts.

In this Letter we report the discovery and properties of what appears to be a merger of two $z\sim5$ galaxies, both undergoing intense, recent and apparently coordinated bursts of star formation, that are magnified (and doubly-imaged) by factors of $\sim$12--15 by the massive $z=0.44$ galaxy cluster MACS J0417.5-1154. While evidence for major mergers of starburst galaxies has been previously found at similar redshifts \citep[e.g.,][]{Hashimoto2019,Romano2021,Hsiao2022}, these are for mergers of massive galaxies. In contrast, our two star-bursting galaxies have very low masses of \Mstars~$\sim10^{7}$\Msun\ each. This system is thus one of the lowest-mass galaxies ever observed at high-$z$, and almost certainly the lowest-mass major merger triggering elevated SF activity known outside of the local Universe.

Throughout this Letter we assume the $\Omega_{\rm M}=0.3$, $\Omega_\Lambda=0.7$, $H_0=70$~${\rm km \: s^{-1} \: Mpc^{-1}}$ cosmology and the \citet{Chabrier2003} IMF.

\section{The lensed emission line system at $z\sim5$}
\label{sec:discovery}
This work utilises NIRCam imaging observations of the MACS J0417.5-1154 cluster field obtained in the Canadian NIRISS Unbiased Cluster Survey (CANUCS, \citealt{willott2022}). CANUCS observes each of its cluster fields in the NIRCam filters F090W, F115W, F150W, F200W, F277W, F356W, F410M and F444W with exposure times of 6.4\,ks each, reaching S/N between 5 and 10 for a AB=29 point source. 
We also utilized archival data of HST/ACS imaging observations with F435W, F606W, and F814W filters (HST-GO-11103 PI Ebeling, HST-GO-12009 PI von der Linden, HST-GO-14096 PI Coe, HST-GO-16667 PI Brada\v{c}). 
We searched the CANUCS  NIRCam images for F410M medium-band-excess line emitters tagged by their very blue F410M-F444W colour. 
Visual inspection of this larger sample confirmed a number of color-excess objects, including several close galaxy pairs, but flagged the present lensed system as of special interest because of its appearance as a close pair of galaxies (ELG1 and ELG2) that is clearly doubly-lensed, for a total of four components (ELG1.1, ELG2.1, ELG1.2, and ELG2.2; see Fig.~\ref{fig:RGB_SED}).

Our image processing and photometry procedures are similar to those we described in \citet{Noirot_2022} for the SMACS 0723 field. Aperture photometry for the four components, plotted in Fig.~\ref{fig:RGB_SED} and listed in online Table~S1, was measured in 0.3"-diameter apertures on HST/ACS and JWST/NIRCam images that were PSF-homogenized to the resolution of the F444W data.

In each one of the four components, photometry shows clear excess in both F410M and F277W, and drops out in F606W, indicating that we are seeing {\sc [Oiii]}+H$\beta$ and H$\alpha$ emission from $5.0<z<5.2$.
The symmetric appearance and matching colors of the two images (color panel in Fig.~\ref{fig:RGB_SED}) confirm the ELG1.1-ELG2.1 system is the counterpart of the ELG1.2-ELG2.2 system.
We update the \citet{Mahler2019} lensing model using the ELG1.1-ELG1.2 image pair as an additional constraint, and find the $z=5.05$ critical line (white dashed line in Fig.~\ref{fig:RGB_SED}) and magnification factors $\mu\sim$~12--15 (Table~\ref{tab:phot_phys}). The separations between the two components in both images are measured by projecting their positions into the source plane, giving a separation of $\sim 1.2$ kpc, consistent for both images (North image: 0.198\arcsec; South image: 0.202\arcsec).

%% Table w/o photometry
\begin{table*}
    \centering
    \caption{Inferred properties of the four elements of the lensed galaxy pair.  
    %All quantities are measured in this work except for the $\mu$ values which are from the \citet{Mahler2019} customised model.
    All the physical properties are obtained from 0.3" aperture photometry after applying aperture and lensing corrections. The $\mu$ values are from our model. The two rightmost columns are results for combined photometry from the two images.
  	}
    \label{tab:phot_phys}
    \begin{tabular}{lccccccc} % four columns, alignment for each
		\hline\hline
		Property & units & ELG1.1 & ELG2.1 & ELG1.2 & ELG2.2 & ELG1 (combined) & ELG2 (combined)\\
		\hline
		$\mu$  & &  $14.1^{+0.6}_{-0.4}$ & $16.0^{+0.5}_{-0.8}$ & $12.6^{+0.3}_{-0.4}$ & $13.5^{+0.4}_{-0.5}$ & $26.7^{+0.9}_{-0.8}$ & $29.5^{+0.9}_{-1.3}$ \\
		\hline
		$\log(M_\star/M_\odot)$ & & $6.88^{+0.15}_{-0.05}$ & $6.84^{+0.31}_{-0.25}$ & $6.96^{+0.36}_{-0.08}$ & $6.95^{+0.35}_{-0.40}$ & $6.98^{+0.30}_{-0.14}$ & $6.81^{+0.41}_{-0.50}$ \rule[-2mm]{0mm}{4mm} \\
		SFR & \Msun\ ${\rm yr}^{-1}$ & $0.23^{+0.00}_{-0.04}$ & $0.12^{+0.06}_{-0.11}$ & $0.22^{+0.05}_{-0.13}$ & $0.16^{+0.07}_{-0.16}$ & $0.23^{+0.02}_{-0.03}$ & $0.11^{+0.10}_{-0.11}$ \rule[-2mm]{0mm}{4mm} \\
		$\log({\rm sSFR/Gyr^{-1}})$ & & $1.49^{+0.15}_{-0.14}$ & $1.27^{+0.46}_{-1.55}$ & $1.39^{+0.45}_{-0.45}$ & $1.26^{+0.50}_{-2.81}$ & $1.39^{+0.32}_{-0.21}$ & $1.25^{+0.67}_{-3.27}$ \rule[-2mm]{0mm}{4mm} \\
		$A_{V}$ & mag & $0.21^{+0.03}_{-0.05}$ & $0.21^{+0.47}_{-0.14}$ & $0.21^{+0.50}_{-0.09}$ & $0.24^{+0.78}_{-0.12}$ & $0.24^{+0.05}_{-0.08}$ & $0.27^{+0.95}_{-0.14}$ \rule[-2mm]{0mm}{4mm} \\
		$Z_{\rm SED}$ & $Z_\odot$ & $0.18^{+0.02}_{-0.03}$ & $0.15^{+0.05}_{-0.07}$ & $0.15^{+0.05}_{-0.08}$ & $0.15^{+0.12}_{-0.08}$ & $0.18^{+0.02}_{-0.08}$ & $0.15^{+0.29}_{-0.09}$ \rule[-2mm]{0mm}{4mm} \\
		\hline
		H$\alpha$ flux & $10^{-19}\ {\rm erg\ s^{-1}\ cm^{-2}}$ &  2.95$\pm$0.40 & 1.79$\pm$0.35 & 2.88$\pm$0.49 & 2.61$\pm$0.46 & 2.92$\pm$0.44 & 2.16$\pm$0.40 \\
		H$\beta$ flux & $10^{-20}\ {\rm erg\ s^{-1}\ cm^{-2}}$ &  10.34$\pm$1.40 & 6.26$\pm$1.24 & 10.12$\pm$1.7 & 9.15$\pm$1.6 & 10.24$\pm$1.56 & 7.59$\pm$1.41 \\
		{\sc [Oiii]} flux & $10^{-19}\ {\rm erg\ s^{-1}\ cm^{-2}}$ &  9.78$\pm$1.08 & 6.28$\pm$0.93 & 10.22$\pm$1.4 & 7.32$\pm$1.29 & 9.99$\pm$1.22 & 6.76$\pm$1.10 \\
		$\log$ EW$_{\rm rest}$ (H$\alpha$) &  & $3.16^{+0.07}_{-0.08}$ & $3.04^{+0.08}_{-0.10}$ & $3.11^{+0.08}_{-0.09}$ & $3.07^{+0.08}_{-0.09}$ & $3.14^{+0.07}_{-0.09}$ & $3.05^{+0.08}_{-0.10}$ \rule[-2mm]{0mm}{4mm} \\
		$\log$ EW$_{\rm rest}$ ({\sc [Oiii]+}H$\beta$) & & $3.38^{+0.06}_{-0.06}$ & $3.28^{+0.07}_{-0.08}$ & $3.37^{+0.07}_{-0.08}$ & $3.23^{+0.07}_{-0.08}$ & $3.37^{+0.06}_{-0.07}$ & $3.25^{+0.07}_{-0.08}$ \rule[-2mm]{0mm}{4mm} \\
		R3 & & 7.08$\pm$1.23 & 7.51$\pm$1.86 & 7.56$\pm$1.65 &  5.99$\pm$1.49 &  7.31$\pm$1.43 &  6.67$\pm$1.64 \rule[-2mm]{0mm}{4mm} \\
		$Z_{\rm R3}$ & $Z_\odot$ &  $0.23^{+0.00}_{-0.13}$ & $0.23^{+0.00}_{-0.14}$ & $0.23^{+0.00}_{-0.13}$ & $0.11^{+0.13}_{-0.05}$ & $0.23^{+0.00}_{-0.13}$ & $0.16^{+0.07}_{-0.09}$ \rule[-2mm]{0mm}{4mm} \\
		\hline
	\end{tabular}
\end{table*}

\section{Methodology}\label{sec:method}
\subsection{SED fitting}\label{subsec:sed}
We used the {\tt dense basis} SED-fitting method \citep{Iyer+17,Iyer+19} to determine the stellar masses, specific star formation rates (sSFRs), metallicities, dust extinction values, and non-parametric star formation histories (SFHs) from the broad/medium band photometry of the four components. 
We stress that {\tt dense basis} does not assume a parametric SFH; rather, it infers the SFH in a flexible way from the data.
We ran {\tt dense basis} in its default configuration, including the 
\citet{Calzetti2000} dust law and \citet{Chabrier2003} IMF; redshifts were  allowed within $z=0$--$16$, and we assumed an exponential $A_V$ prior with scale value of $A_{V}=3.0$ mag.  This high dust prior allows extremely dusty solutions, important for discriminating between high-$z$ sources and low-$z$ dusty interlopers. 
Note that, in {\tt dense basis}, nebular emission lines are included in the model spectrum in a self-consistent way with the metallicity, and we fitted to the photometry including F277W, F410M and F444W filters where strong emission lines are thought to fall.

To obtain intrinsic values of the physical parameters, before SED-fitting we applied aperture and lensing magnification corrections to the photometry in online Table S1. Lensing corrections were obtained from the lens model (see Sec.~\ref{sec:discovery}), and we applied aperture flux corrections of 2$\times$, consistent with the growth curves measured using elliptical apertures on the NIRCam images.
The SED-fitting results are shown in Figure \ref{fig:RGB_SED} and in Table~\ref{tab:phot_phys}.

\subsection{Emission line maps}\label{subsec:eline}

\begin{figure*} 
	\includegraphics[width=2\columnwidth]{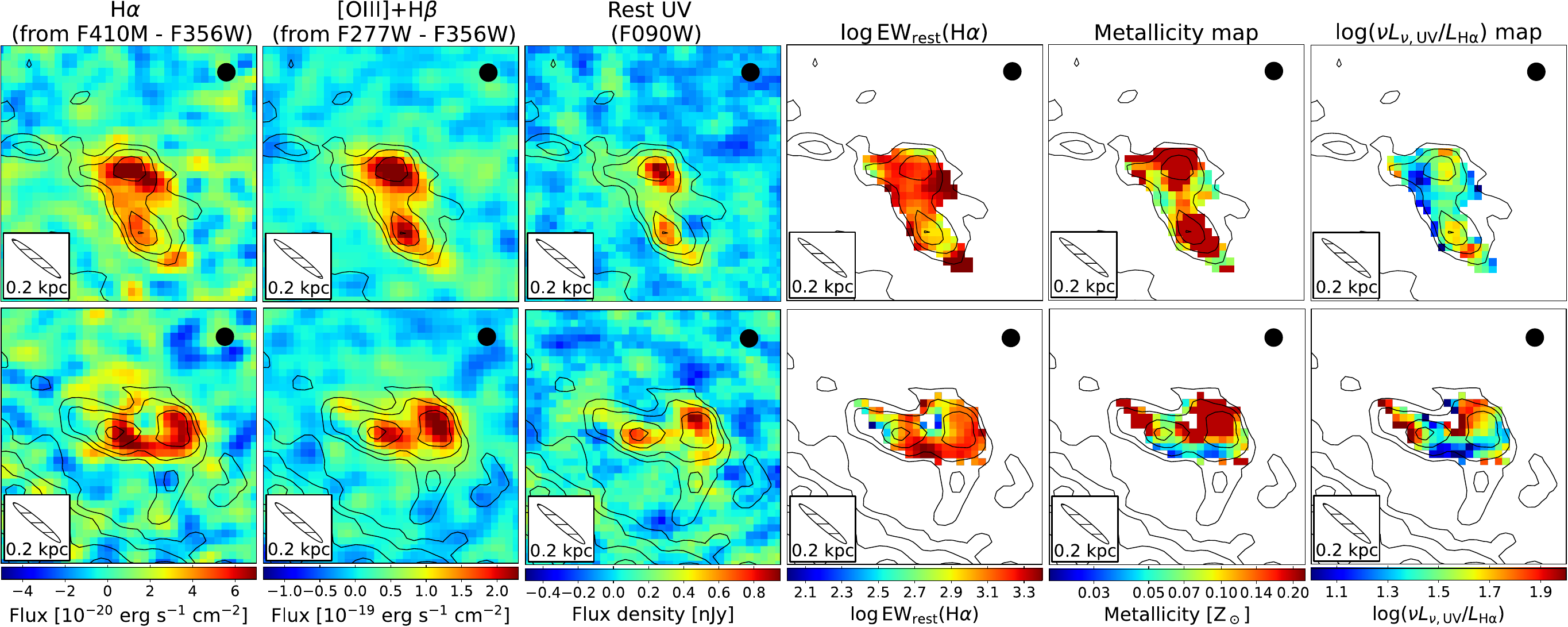}
    \caption{Property maps for both images of the galaxy pair, convolved to F444W spatial resolution. The panels are 1.4" on the side, with 0.04" per pixel. The black dot at the top right of each panel shows the PSF FWHM in the image plane. The hatched ellipses at the bottom left represent the lensed image of a circle whose diameter is 0.2 physical kpc in the source plane, giving an illustration of the physical scale -- which is anisotropic in the image plane -- and an insight into the physical separation of the two components. The delensed physical separation of ELG1 and ELG2 cores is $\sim1.2$ kpc measured directly in the source plane. In all panels the black contours are linear flux in F356W and show F356W continuum, which corresponds to rest-frame $\sim$5800\AA\ (i.e., stellar) continuum at $z=5.1$. The fluxes are not lens corrected. 
    }
    \label{fig:cutouts}
\end{figure*}

The \OIIIHB\ and the \Ha\ emission lines boost the F277W and F410M fluxes, respectively, but the F356W photometry is free from strong emission lines and traces rest-frame optical stellar continuum.  Thus, subtracting the F356W image from F277W (or F410M) gives the \OIIIHB\ (and \Ha) emission line maps. The difference images using PSF-matched images are shown in Fig.~\ref{fig:cutouts} with F356W emission (i.e., stellar continuum) overplotted as black contours for comparison.

To quantitatively evaluate the emission line flux maps we calibrated the difference images of Fig.~\ref{fig:cutouts} as follows. The observed flux density in filter $i$ ($f_{\nu,i}$) where an emission line or lines fall ($i=$F277W or F410M in this work) can be written as
\begin{equation}\label{eq:line-flux-density}
    f_{\nu,i} = \frac{\lambda_i^2}{c}\frac{F_{\rm line} + f_{\lambda,i}\Delta\lambda_i}{\Delta\lambda_i},
\end{equation}
where $\lambda_i$ is the pivot wavelength of filter $i$, $\Delta \lambda_i$ is the bandwidth of filter $i$, $F_{\rm line}$ is the line flux that falls into the filter, and $f_{\lambda,i}$ is the flux density of the continuum level in the filter.
If a filter $j$ is free from strong emission lines, Eq.~\ref{eq:line-flux-density} reduces to
\begin{equation}
    f_{\nu,j} = \frac{\lambda_j^2}{c}f_{\lambda,j}.
\end{equation}
Therefore, if we assume a continuum that is flat in $f_\nu$ between filters $i$ and $j$ (i.e., $\lambda_i^2f_{\lambda,i} = \lambda_j^2f_{\lambda,j}$), the emission line flux $F_{\rm line}$ can be calculated through the difference $f_{\nu,i} - f_{\nu,j}$ as
\begin{equation}\label{eqn:Eflux}
    F_{\rm line} \approx \frac{c\Delta\lambda_{i}}{\lambda_{i}^2}(f_{\nu,i} - f_{\nu,j}).
\end{equation}
Similarly, the ratio of $f_{\nu,i}$ and $f_{\nu,j}$ gives the observed equivalent width (EW$_{\rm 
obs}$) of the line under the same assumption as
\begin{equation}\label{eqn:EW}
    {\rm EW_{\rm obs}} \approx \Delta\lambda_i \left(\frac{f_{\nu,i}}{f_{\nu,j}} - 1 \right).
\end{equation}
Using Eqn.~(\ref{eqn:Eflux}), (\ref{eqn:EW}), and F356W as the reference filter for the continuum level ($j=$F356W), we calibrated the H$\alpha$ and {\sc [Oiii]}+H$\beta$ emission line flux maps and the EW maps in Fig.~\ref{fig:cutouts}.

Furthermore, assuming the intrinsic line flux ratio  H$\alpha$/H$\beta=2.85$ and {\sc [Oiii]5007}/{\sc [Oiii]4959} $=2.98$, we made the {\sc [Oiii]}-to-H$\beta$ ratio (${\rm R3 = } ${\sc [Oiii]}${\rm 5007/H\beta}$) map.
R3 is one of the indicators of gas-phase metallicity and so this map can be converted into a metallicity map \citep[e.g.,][]{Nakajima+22}. R3 can have values that are degenerate with metallicity, but since the two galaxies are extremely low-mass and early in a major SF episode (Sec.~\ref{subsec:phys_interp}), we adopted the lower metallicity solution. In some pixels around the galaxy centers (Fig.~\ref{fig:cutouts}), the R3 values slightly exceed the maximum value predicted by the R3-metallicity relation, and in these pixels we set the metallicity to the value that maximizes R3 ($Z=0.2Z_\odot$). We have not applied dust corrections, but the resulting line ratio and metallicity do not change significantly if we assume $A_V=0.2$ mag indicated by SED fitting. Also, we ignore the contribution of {\sc [Nii]} emission to the F410M flux density, since our high {\sc [Oiii]}/H$\beta$ ratio ($\sim7$), implies that contribution is $\lesssim5\%$ of H$\alpha$ \citep[BPT diagram;][]{Veilleux1987}.
The presence of AGN would affect the line ratio and subsequent conclusions, but we do not think AGNs are present given the low masses of the galaxies; however, to be sure will require spectroscopy.

Also of interest is the rest-frame UV to H$\alpha$ luminosity ratio. Since the H$\alpha$ emission line is sensitive to SF activity on shorter time-scales than is rest-frame UV luminosity, this ratio is often used as a burstiness parameter \citep[e.g.,][]{faisst_recent_2019}. We thus show in Fig.~\ref{fig:cutouts} the
$\nu L_{\nu,{\rm UV}}/L_{\rm H\alpha}$ map made from the H$\alpha$ emission line flux map and the F090W image. 
Again, assuming $A_V=0.2$ mag increases $\nu L_{\nu,{\rm UV}}/L_{\rm H\alpha}$ by 0.14 dex but does not change the following results.

We also followed the above procedure but using the F277W, F356W, and F410M aperture photometry to derive the spatially-integrated line fluxes, EWs, R3 values, and metallicities of the four components, and the results are shown in the bottom of Table~\ref{tab:phot_phys}. 

\section{Results and Discussions}
\subsection{A likely merger at $z\sim5$}
\label{subsec:phys_interp}

Given that both ELG1 and ELG2 are expected to be within the narrow redshift range $5.0<z<5.2$ with very small projected separation ($\sim1.2$ physical kpc), it is highly likely ELG1 and ELG2 are physically associated. Such a close pair can be interpreted as merging galaxies or star-forming clumps within a single galaxy. In the case of ELG1-ELG2, the former is more likely. 

First, the \Ha\ image shows a bridge-like structure connecting the two cores (see left-most panel in Fig.~\ref{fig:cutouts}), while the rest-frame optical continuum (F356W image) may contain a tidal feature in the south of ELG1.2 and east of 1.1 (e.g., see black contours in Fig.~\ref{fig:cutouts}). While not uniquely so, such morphological features are consistent with a merger. 
Second, we see no evidence for an underlying, larger galaxy containing the two clumps in the rest-frame optical continuum at $\lambda_{\rm rest}\sim5800$ \AA\ (F356W). The age of the Universe at $z\sim5.1$ is comparable to or shorter than the life-time of the A stars that dominate at $\lambda_{\rm rest}\sim5800$ \AA, and thus the F356W image should well delineate every stellar component. ELG1 and ELG2 are distinct in both F356W images (see Fig.~\ref{fig:RGB_SED} and black contours in Fig.~\ref{fig:cutouts}) and are also coincident only with regions of line emission, which indicates stars are distributed only in the SF regions.
Dust reddening might help to hide the underlying galaxy, but an extremely high dust opacity is needed to make the 5800 \AA\ continuum invisible, which seems unlikely.
Finally, the SFHs for ELG1 and ELG2 show a synchronized start of their SF bursts (Fig.~\ref{fig:RGB_SED}). Such synchronization can be expected from merger-induced star formation, but star-forming clumps could be expected to burst asynchronously;  they could also come in multiplicities higher than two. 
Considering these facts, the ELG1+ELG2 system is likely to be a merger system rather than SF clumps within a single galaxy, although we cannot rule out the possibility of SF clumps completely.
%A spectroscopic follow-up observation might make this sure.

\subsection{Physical properties of the system}
\label{subsec:merger-properties}

SED fitting (Fig.~\ref{fig:RGB_SED} and Table~\ref{tab:phot_phys}) reveals that the delensed stellar masses of the two galaxies are both $\log(M_\star/M_\odot)\sim6.9$.
The low $\nu L_{\nu,{\rm UV}}/L_{\rm H\alpha}$ ratio can only be explained by a bursty SFH \citep[e.g.,][]{Mehta2022}.
Their model SFHs indeed show recently-started ($\lesssim$ 100 Myr) SF bursts, consistent with the large \Ha\ EWs as well. These results suggest the ELG1-ELG2 system is an equal-mass merger of ultra-low-mass galaxies, with interaction-induced SF bursts, that we are witnessing at its early stages. 

The \Ha\ morphology (Fig.~\ref{fig:cutouts}) shows a bridge-like structure connecting the galaxy centers in both images.  In contrast, the \OIIIHB\ image shows line emission that peaks at the center of the stellar component and the bridge-like structure is not apparent; the \OIIIHB\ morphology is similar to that of the rest-frame UV flux map. Given that H$\alpha$ emission marks SF activity happening on shorter timescales than rest-frame UV luminosity, these facts suggest that the ELG1-ELG2 system has a bridge-like structure showing {\it in-situ} star formation, possibly induced by the disruptive forces of the merger event.

Figure \ref{fig:cutouts} also shows the EW, metallicity, and  $\nu L_{\nu,{\rm UV}}/L_{\rm H\alpha}$ maps.  At the galaxy centers, the rest-frame H$\alpha$ EW is $\log({\rm EW})\sim3.0$, metallicity is $Z\sim0.2Z_\odot$, and the UV-to-H$\alpha$ ratio is $\log(\nu L_{\nu,{\rm UV}}/L_{\rm H\alpha})\sim1.6$, whereas in the bridge region, the H$\alpha$ EW is slightly higher ($\log({\rm EW})\sim3.2$), the metallicity is lower ($Z\sim0.08Z_\odot$), and the UV-to-H$\alpha$ ratio is lower as well ($\log(\nu L_{\nu,{\rm UV}}/L_{\rm H\alpha})\lesssim1.4$).  What do we think this means? 

First, both the rest-frame H$\alpha$ EW and UV-to-H$\alpha$ ratio are probes of the age of the SF burst \citep[e.g.,][]{inoue_rest-frame_2011,Mehta2022}, and the ages predicted from the EW(H$\alpha$) and $\nu L_{\nu,{\rm UV}}/L_{\rm H\alpha}$ are also $\lesssim100$ Myr, consistent with the SED fitting results (c.f., Section \ref{subsec:sed}). Notably, the high EW(H$\alpha$) and low $\nu L_{\nu,{\rm UV}}/L_{\rm H\alpha}$ in the bridge suggest that the SF activity there has been induced only very recently. 
Alternatively the low $\nu L_{\nu,{\rm UV}}/L_{\rm H\alpha}$ in the bridge may indicate ISM shocks, although either scenario is consistent with a merger.

Second, the metallicity map shows the galaxy centres are already metal-enriched to $Z\sim0.2Z_\odot$ even though the galaxies have very low masses ($\lesssim10^7\ M_\odot$).
The R3-based metallicity values in the maps are in good agreement with those from SED fitting of 0.3" aperture photometry (Table \ref{tab:phot_phys}), which are centred on ELG1 and ELG2 image centroids and so measure properties of the galaxy centres.  The metallicity map also show a possible lower metallicity ($Z\sim0.08Z_\odot$) in the bridge as compared to the galaxy centres.
Assuming this system is a major merger, this could be evidence for metallicity gradients in high-$z$ galaxies as the low-$Z$ bridge could have formed from stripped material from the outskirts of one of the merging galaxies. 

In summary, we interpret ELG1+ELG2 as an equal-mass merger of two ultra-low-mass galaxies at 
$z\sim5.1$. Their interaction recently ($<$100~Myr) induced intensive, simultaneous, and ongoing bursts of SF and formed a bridge of star-forming material and possibly also a tidal tail. This bridge is undergoing a particularly recent star-forming burst with metallicity that's lower still than that at the galaxy centres. 

\subsection{The assembly of a low-mass galaxy at $z\sim5$}
\label{subsec:assembly-of-mass}

The {\tt dense basis} SFHs of ELG-1 and ELG-2 (Fig.~\ref{fig:cutouts}) show both galaxies are undergoing a $\lesssim100$~Myr-old burst of star formation. About 35\% of the observed mass of each galaxy was formed during this burst, while $\sim$65\% formed before it. This means that the two galaxies had \Mstars~$\sim5\times10^6\ M_\odot$ each before the burst, and the formation of a combined $\sim6\times10^6$\Msun\ was induced by the starburst.

The burst is ongoing, so it is difficult to know how many additional stars will form as it continues, but we can make crude estimates: On the lower end, (Case 1) we can assume that the burst has peaked and SFRs will decline symmetrically in time. This will produce a further $\sim6\times10^6\ M_\odot$, so that once ELG-1 and ELG-2 merge, the resulting galaxy will have \Mstars~$\sim2\times10^7$\Msun, $\sim$50\% formed during the burst, and 25\% formed pre-burst in each of the original galaxies.  

The timescales of equal-mass merger can be a few times the dynamical time, $t_{\rm dyn}$  \citep[e.g.,][]{McCavana2012,Solanes2018}. Therefore (Case 2), if instead of declining as in Case~1, the SFRs hold at their current levels for 100~Myr ($\sim2.5 t_{\rm dyn}$) the merged galaxy will have \Mstars~$\sim5\times10^7\ M_\odot$, i.e.\ 10$\times$ the pre-interaction mass of ELG-1 or ELG-2. 
Of this final mass, 80\% will be new stars formed in the burst.  Finally (Case~3), if the SFHs do not flatten or decline but continue to increase, the post-merger galaxy will have potentially formed even more stars during the burst.  

In conclusion, the galaxy that forms after the burst will have a mass that's at least 4$\times$ the mass of ELG-1 or ELG-2, and will have formed at least half of that mass during the ongoing SF burst. 
This conclusion holds even if ELG-1 and ELG-2 are SF clumps in a (low mass) galaxy and not a merger. If the system is a merger, and if merger-induced starbursts are common in high-$z$ low-mass galaxies, then a large fraction of the high-$z$ cosmic stellar mass density may have formed due to galaxy-galaxy interactions.

\subsection{Contribution to keeping the Universe ionized}
\label{subsec:reionization}

The formation of stars in a galaxy is accompanied by the production of rest-frame UV photons from short-lived massive stars. Thus, we can estimate the total number of ionizing photons, $N_{\gamma_{\rm ion}}$, produced by the system up to time $t$ over the course of its evolution:
$N_{\gamma_{\rm ion}}(t) = \int_0^{t}\dot{N}_{\gamma_{\rm ion}}dt \propto \int_0^t L_{\rm UV} dt \propto \int_0^t {\rm SFR} dt\propto M_\star(t)$.  I.e., the number of ionizing photons produced is directly proportional to the stellar mass made. If these photons escape the system, they will be available for reionizing the Universe in the Epoch of Reionization (EoR) and keeping it ionized thereafter. 

Taking the \Mstars\ estimates from Sec.~\ref{subsec:assembly-of-mass}, we expect that at least the same number of the ionizing photons will be produced within the shortest timescale burst (Case 1 in Sec.~\ref{subsec:assembly-of-mass}) as the two galaxies together have made before the burst. In longer-lasting bursts, the fraction of ionizing photon produced during the burst can be 80\% (Case 2) or more (Case 3) of the total produced since $t=0$. In terms of the production rate ($\dot{N}_{\gamma_{\rm ion}}$), since the current SFRs during the burst are  $\sim10\times$ those before the burst (see SFHs in Fig.~\ref{fig:RGB_SED}), the $\dot{N}_{\gamma_ {\rm ion}}$ in the two galaxies are also $\sim10$ times higher than before the burst. 

Furthermore, we calculate ionizing photon production efficiency ($\xi_{\rm ion}$) from $L_{\rm H\alpha}$ and $L_{\nu, {\rm UV}}$ assuming case B recombination \citep{Leitherer1995} and zero escape fraction, and obtain $\log(\xi_{\rm ion}/{\rm Hz\ erg^{-1}})=25.64^{+0.21}_{-0.23}$ at the galaxy centres. This value is higher than the canonical values \citep[e.g.,][]{robertson_new_2013,matthee_production_2017} or LBGs at similar redshifts \citep{bouwens_lyman-continuum_2016}, and is comparable to other extreme emission-line galaxies at higher or lower redshifts \citep{tang_mmtmmirs_2019,Matthee2022}. In the bridge region, the $\xi_{\rm ion}$ value is even higher, $\log(\xi_{\rm ion}/{\rm Hz\ erg^{-1}})=25.80^{+0.20}_{-0.10}$, and is one of the highest values in the high-$z$ universe.
Note that, similar to $\nu L_{\nu, {\rm UV}}/L_{\rm H\alpha}$, assuming $A_V=0.2$ mag instead reduces $\xi_{\rm ion}$ values by $\sim0.14$ dex, but does not change the conclusion here.

These arguments suggest that at least as many ionizing photons are produced, with high efficiency $\xi_{\rm ion}$, during the (possibly merger-induced) starburst as in the entire pre-burst history of the two galaxies. Notably, galaxy interactions are also thought to help ionizing photons to escape \citep[e.g.,][]{Rauch2011}. Thus, our results suggest that, if merger-induced SF bursts are common among high-$z$ low mass galaxies, then interacting low-mass galaxies could be a major contributor to the Cosmic ionizing photon budget.

\section{Conclusions}
In this Letter we (1) described a lensed, doubly imaged system of two ultra-low-mass ($5\times10^6$\Msun\ each) galaxies at $z\sim5.1$ undergoing a coordinated young SF burst induced by an apparent equal mass major merger, (2) find that the resulting galaxy is expected to form at least half of its stellar mass during the ongoing SF burst, and (3) that a large number of ionizing photons is produced with a high efficiency during the burst. If ELG1-ELG2 is confirmed as a merger, then mergers may play an important role in the assembly of -- and ionizing photon production by -- low mass galaxies at high redshift.

%%%%%%%%%%%%%%%%%%%%%%%%%%%%%%%%%%%%%%%%%%%%%%%%%%
\section*{Acknowledgements}

% \MNRASinstructions{The Acknowledgements section is not numbered. Here you can thank helpful
% colleagues, acknowledge funding agencies, telescopes and facilities used etc.

We thank the anonymous referee for the useful comments and suggestions.
This research was enabled by grant 18JWST-GTO1 from the Canadian Space Agency, funding from the Natural Sciences and Engineering Research Council of Canada.
YA is supported by a Research Fellowship and Overseas Challenge Program by Japan Society of Promotion of Science. MB acknowledges support from the Slovenian national research agency ARRS through grant N1-0238 and the program HST-GO-16667, provided through a grant from the STScI under NASA contract NAS5-26555.

%%%%%%%%%%%%%%%%%%%%%%%%%%%%%%%%%%%%%%%%%%%%%%%%%%
\section*{Data Availability}

%\MNRASinstructions{The inclusion of a Data Availability Statement is a requirement for articles published in MNRAS. Data Availability Statements provide a standardised format for readers to understand the availability of data underlying the research results described in the article. The statement may refer to original data generated in the course of the study or to third-party data analysed in the article. The statement should describe and provide means of access, where possible, by linking to the data or providing the required accession numbers for the relevant databases or DOIs.}

Raw JWST data used in this work will be available from the {\it Mikulski Archive for
Space Telescopes} (\url{https://archive.stsci.edu}), 
at the end of the 1-year proprietary time. Processed data products will be available on a similar timescale at \url{http://canucs-jwst.com}.



\bibliographystyle{mnras}
\bibliography{reference} % if your bibtex file is called example.bib

%\begin{thebibliography}{99}
%\end{thebibliography}

%%%%%%%%%%%%%%%%%%%%%%%%%%%%%%%%%%%%%%%%%%%%%%%%%%

%%%%%%%%%%%%%%%%% APPENDICES %%%%%%%%%%%%%%%%%%%%%

%\newpage

%%%%%%%%%%%%%%%%%%%%%%%%%%%%%%%%%%%%%%%%%%%%%%%%%%

% Don't change these lines
\bsp	% typesetting comment
\label{lastpage}
\end{document}